\documentclass[letterpaper,10pt,twocolumn]{article}
\usepackage{ol2}
\usepackage[latin1]{inputenc}
\usepackage[T1]{fontenc}
\usepackage{amstext,upgreek}
\newcommand{\units}[1]{\;\mathrm{#1}}
\newcommand{\txt}[1]{{\text{\scriptsize #1}}}
\begin{document}
\twocolumn[%
\title{Excitation mapping of whispering gallery modes in silica microcavities}
\author{G. Lin,$^{1,2}$ B. Qian,$^{1,3}$ F. Oru\v{c}evi\'{c},$^{1,4}$ Y. Candela,${^1}$ J.-B.~Jager,$^{5}$ Z.~Cai,$^{2}$ V.~Lefèvre-Seguin,${^1}$ J.~Hare,$^{1,*}$}
\address{
$^1$Laboratoire Kastler Brossel, ENS,  UPMC-Paris 6, CNRS --- 24 rue Lhomond, 75231 Paris cedex 05, France\\
$^2$ Department of Physics, Xiamen University, Xiamen 361005, Fujian, P. R. China\\
$^3$ Nanoscience Laboratory, Department of Physics, University of Trento, Via Sommarive 14, I-38100 Povo, Italy\\
$^4$ Department of Physics and Astronomy, University of Sussex, Falmer, Brighton, BN1 9QH, United Kingdom\\
$^5$ CEA Grenoble DRFMC/SP2M/SINAPS Minatec, 17 rue des Martyrs 38054 Grenoble cedex 9, France\\
$^*$Corresponding author: Jean.Hare@lkb.ens.fr}

\begin{abstract}
We report the direct observation of the electromagnetic-field distribution of whispering--gallery modes in silica microcavities (spheres and toroids). It is revealed by their excitation efficiency with a tapered fiber coupler swept along the meridian. The originality of this method lies in the use of the coupler itself for the near field mapping, eliminating the need of additional tools used in previous work. This method is successfully applied to  microspheres and microtoroids.
\end{abstract}

] 

Silica optical microcavities formed by surface tension, like microspheres\cite{KlitzingLong01} and more recently microtoroids~\cite{ArmaniKippenberg2003}, feature whispering gallery modes (WGM) that have drawn a large interest in the last two decades. WGMs are characterized by ultra-high quality factors (up to $10^{10}$ in the infrared) and moderate mode volumes (down to about $100\;\lambda^3$ for a typical diameter of $50\units{\upmu m}$). They have led to many applications in various fields like CQED, microlasers, telecommunications and biosensing. A WGM can be seen as a nearly grazing ray guided around the resonator by successive total internal reflections and fulfilling a phase matching condition after one roundtrip. The ``fundamental mode'', corresponding to a ray tightly bound close to the equator of the cavity, is characterized by a single antinode along both the polar and radial directions, and achieves the smallest mode volume. Most applications depending on the mode volume can be optimized by selectively working with this mode, which needs to be unambiguously identified.

Several papers \cite{GorodetskyIlchenko94,KnightDubreuil95,GotzingerDemmerer2001,DongXiao2008} have been devoted to this question.  In \cite{GorodetskyIlchenko94}, the near-field of a  microsphere is imaged on a camera through a coupling prism and a microscope. Refs.~\cite{KnightDubreuil95} and \cite{GotzingerDemmerer2001}  are based on direct detection of the near-field: a fiber tip is scanned along the sphere surface in order to map the evanescent field, for a fixed excitation frequency. In \cite{DongXiao2008}, the near-field of poorly confined leaky modes was directly imaged on a camera.
The dependance of the coupling efficiency with respect to the mode order that is at the heart of the present paper has been used in \cite{SavchenkovMatsko2005} to filter out high order WGMs of a cm-sized microdisk using an auxiliary coupling prism.

In this paper, we show that near-field mapping does not require the use a fiber tip or coupling prism but can readily bee obtained  using the tapered fiber coupler technique \cite{KnightCheung98,CaiPainter00-prl} and a widely tunable laser. The near-field distribution is revealed through the spatial dependance of the excitation efficiency. We first demonstrate this method on microspheres which have a well known WGM spectrum and then apply it to a microtoroid, for which no simple analytic description exists.

\begin{figure}[b]
  \centering
  \includegraphics[width=50mm]{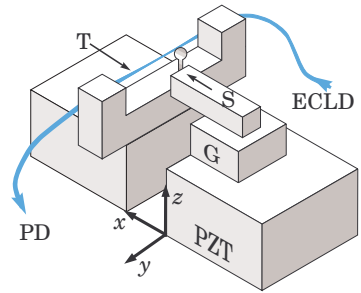}
  \caption{\small Sketch of the experimental setup, with definition of the $x,y,z$ axes used hereafter. S: Sphere (not to scale), T: Taper, PZT: Three-axis piezoelectric stage, G: Goniometer, ECLD: Extended cavity laser diode, PD: Photodiode.}\label{f:setup}
\end{figure}
Beside their polarisation, the WGMs of a microsphere are characterized by three integer orders $n$, $l$ and $m$. The radial order $n$ given by the number of antinodes of the radial function is not considered here. The orders $\ell$ and $m$ are the usual angular momentum quantification numbers, with $m\in[-\ell,\ell]$, corresponding to a field distribution approximately proportional to the spherical harmonic $Y_{\ell}^m$. The fundamental mode then corresponds to the WGM of highest azimuthal number $m=\ell$, which has a single antinode in the polar direction. It is well known\cite{LaiLeung90-1} that a small ellipticity $e$ breaks the spherical symmetry and leads to a frequency shift:
\begin{equation}\label{e:splitting}
\frac{\delta\nu_{\ell,m}}{\nu_{\ell,m}} = -\frac{e}{6}\;\left(1-\frac{3\,m^2}{\ell(\ell+1)}\right)
\approx \frac{e}{3} - \frac{e}{\ell}(\ell-|m|) \ .
\end{equation}
This gives a multiplet of nearly equally spaced modes for $|m|\approx \ell$ with an interval $\Delta\nu=e\;\nu_{\ell,m}/\ell\approx e\,\mathit{FSR}$ where $\mathit{FSR}=c/2\pi n_\txt{S} a\sim 1000\units{GHz}$ is the free spectral range of the cavity ($a$: sphere radius, $n_\txt{S}$: refraction index of silica). It moreover indicates that the resonance frequency decreases with $q=\ell-|m|$ for a prolate spheroid ($e>0$). Hence the fundamental mode has the \emph{largest} frequency of the $n,\ell$ multiplet in our experiments with prolate microspheres.

The  experimental setup is sketched in Fig.~\ref{f:setup}. The microsphere is prepared by CO$_2$ laser melting of a tapered fiber, allowing to finely control its size and eccentricity, and to optimize the parallelism of its revolution axis with the mother-fiber axis \cite{KlitzingLong01}. The taper coupler is produced by pulling a single-mode  optical fiber heated by a microtorch, as described in \cite{OrucevicLefevre2007}. A widely tunable mode-hop-free extended cavity laser diode ($\lambda\sim 775\units{nm}$) is coupled into the tapered fiber. When scanning the laser frequency  the WGM resonances are detected by  an absorption dip on the laser output intensity monitored by a photodiode (PD) and a digital oscilloscope. Our setup is designed to maximize the excitation of  low angular order $q$ WGM. The rotation angles around the $x$ and $y$ axes and the position along the $y$- and $z$-axis should first be  optimized with the goniometer and PZT shown in Fig.~1. The air gap $g$ between the sphere and the coupler is then adjusted with the PZT actuator. The gap is set such that the resonances are in the \emph{undercoupled regime}, where the broadening of the resonance linewidth is negligible.

\begin{figure}[tb]
  \centering
\includegraphics[width=70mm]{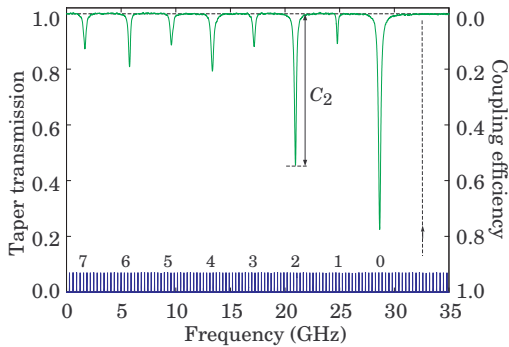}
  \caption{\small Typical taper transmission spectrum for a prolate microsphere of diameter $D\approx 68\units{\upmu m}$ and ellipticity $e\approx0.4\%$. Each line corresponds to the $q=\ell-|m|$ value given below it.  The bottom curve is the spectrum of a confocal Fabry-Perot providing the frequency scale.}\label{f:spectrum}
\end{figure}

Fig.~\ref{f:spectrum} shows an optimized taper transmission spectrum exhibiting a WGM multiplet. The equally spaced lines correspond to  the splitting described by Eq.~\ref{e:splitting}. The missing resonance on the high frequency side at the location pointed by the dotted line ascertain the identification of the $q=0$ WGM, and the subsequent determination of the other $q$ values in this series. The ``excitation'' or ``coupling efficiency'' is defined as the relative depth of this transmission dip $C= 1-T_\txt{res}/T_0$, where $T_0$ is the out-of-resonance transmission and $T_\txt{res}<T_0$ is the on-resonance transmission.

\begin{figure}[th]
  \centering
  \includegraphics[width=70mm]{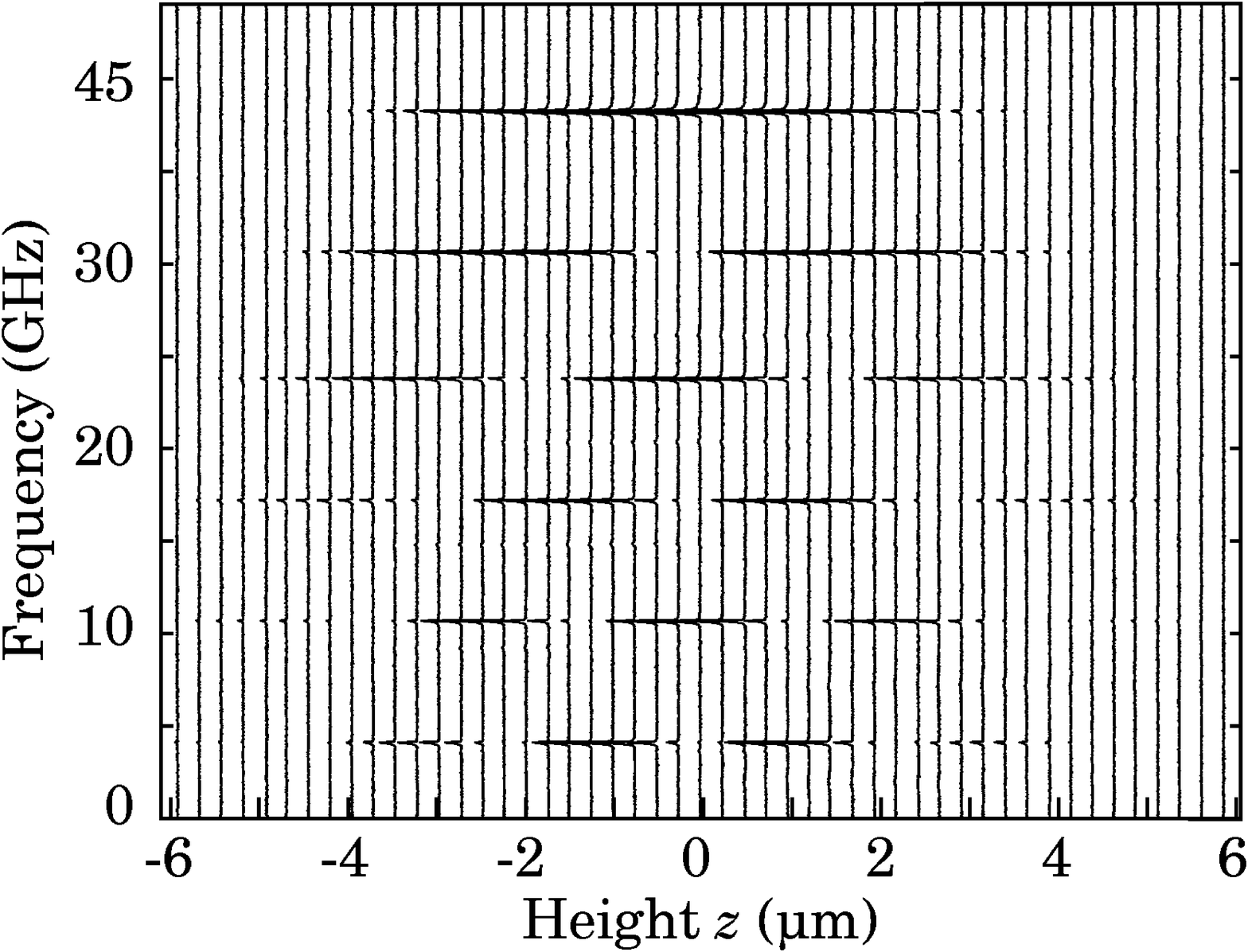}
  \caption{\small Waterfall of transmission spectra obtained for different $z$-positions: each transmission spectrum is plotted at the corresponding $z$ value. }
  \label{f:waterfall}
\end{figure}

We have recorded the WGMs' transmission spectrum for different values of the sphere height $z$. The resulting spectra are displayed as a waterfall in Fig.~\ref{f:waterfall} for a sphere of diameter $2a=56\units{\upmu m}$ and ellipticity $e=1.2\%$. The successive curves are horizontally offset according to the corresponding $z$ value, so that the abscissa corresponds to both the coupling efficiency and the $z$-coordinate. The overlapping absorption dips allow a clear visualization of the field distribution of the different modes, with  $q+1=\ell-|m|+1$ antinodes.

\begin{figure}[h]
  \centering
  \includegraphics[width=70mm]{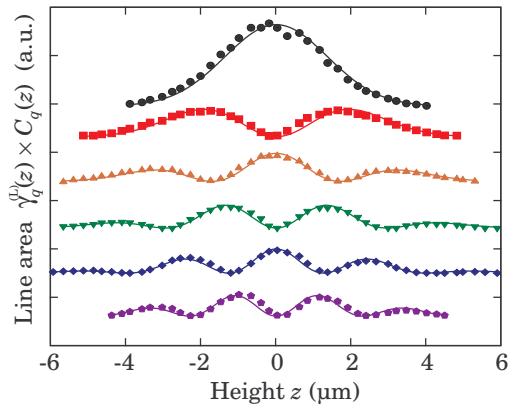}
  \caption{\small Symbols: coupling efficiency as a function of $z$-position, for $q=0\cdots5$, deduced from  Fig.~\protect\ref{f:waterfall}. Solid lines: global fit of the experimental data.}
  \label{f:modes}
\end{figure}

For a detailed analysis of these data we use a simplified expression of the coupling efficiency valid in the thin taper limit, assuming that the overlap integral of the WGM and the coupler fields is simply proportional to the WGM field. In this model, using spherical coordinates $(r,\theta,\phi)$ with origin on the sphere center,  the coupling efficiency writes:
\begin{equation}\label{e:coupleff}
    C_q(r,\theta) = K_q\; \frac{\gamma_q^{(0)}}{\gamma_q^\txt{(L)}}\; \big|Y_\ell^{\ell-q}(\theta)\big|^2 \ \exp(-2\kappa\,(r-a))
\end{equation}
where $K_q$ is a scaling coefficient depending on the taper diameter and its effective index, $\gamma_q^{(0)}$ is the intrinsic linewidth and $\gamma_q^\txt{(L)}$ the observed loaded linewidth, $\kappa^{-1}\approx (n_\txt{S}^2-1)^{-1/2}\; \lambda/2\pi$ is the evanescent wave characteristic depth. The ratio of the two linewidths allows to take into account overcoupling, that is visible on the fundamental mode.

When substituting the spherical harmonic by its  Hermite-Gauss asymptotic expression for large $\ell$, and using a second order expansion of the gap $g=r-a=g_0+z^2/2a$, equation (\ref{e:coupleff}) leads to:
\begin{equation}\label{e:coupleff3}
   \gamma_q^\txt{(L)}(z)  C_q(z) \propto H_q^2\left(\sqrt{\ell}\frac{z}{a}\right)\ \exp\left[-\big( \frac{\ell}{a^2} +\frac{\kappa}{a}  \big)\; z^2\right] \ .
\end{equation}

From the experimental results, we have extracted the coupling efficiency and linewidth of all the resonances, and plotted in Fig.~\ref{f:modes} the normalized area of the resonances, given by the product $C_q(z)×\gamma_q^{(0)}/\gamma_q^\txt{(L)}$. The curves obtained for different $q$ values are offset to evidence the similarity with Fig.~\ref{f:waterfall}.

We then perform a global fit of these curves according to Eq.~\ref{e:coupleff3}, using different amplitudes but the same horizontal scale for the 6 $q$ values. This fit is plotted as solid lines in Fig.~\ref{f:modes} and is in very good agreement with the experimental points, thus proving that our model is  accurate and that the measured profiles are actually related to the field distribution.
A departure from the model are nevertheless observed:  the fitted horizontal $z$-scale does not match the expected one but  is  larger by a factor of about 1.4. This effect does not come from mechanical effects or from the PZT calibration, which can easily be ruled out. It likely results from the non-negligible taper diameter, that is significantly larger than the depth of the evanescent wave $\kappa^{-1}$. This idea has been verified in a simple numerical simulation of the overlap integral, where the finite taper diameter systematically results in an effective $z$ smaller than the taper center position used in the experiment.

\begin{figure}[!h]
  \centering
  \includegraphics[width=70mm]{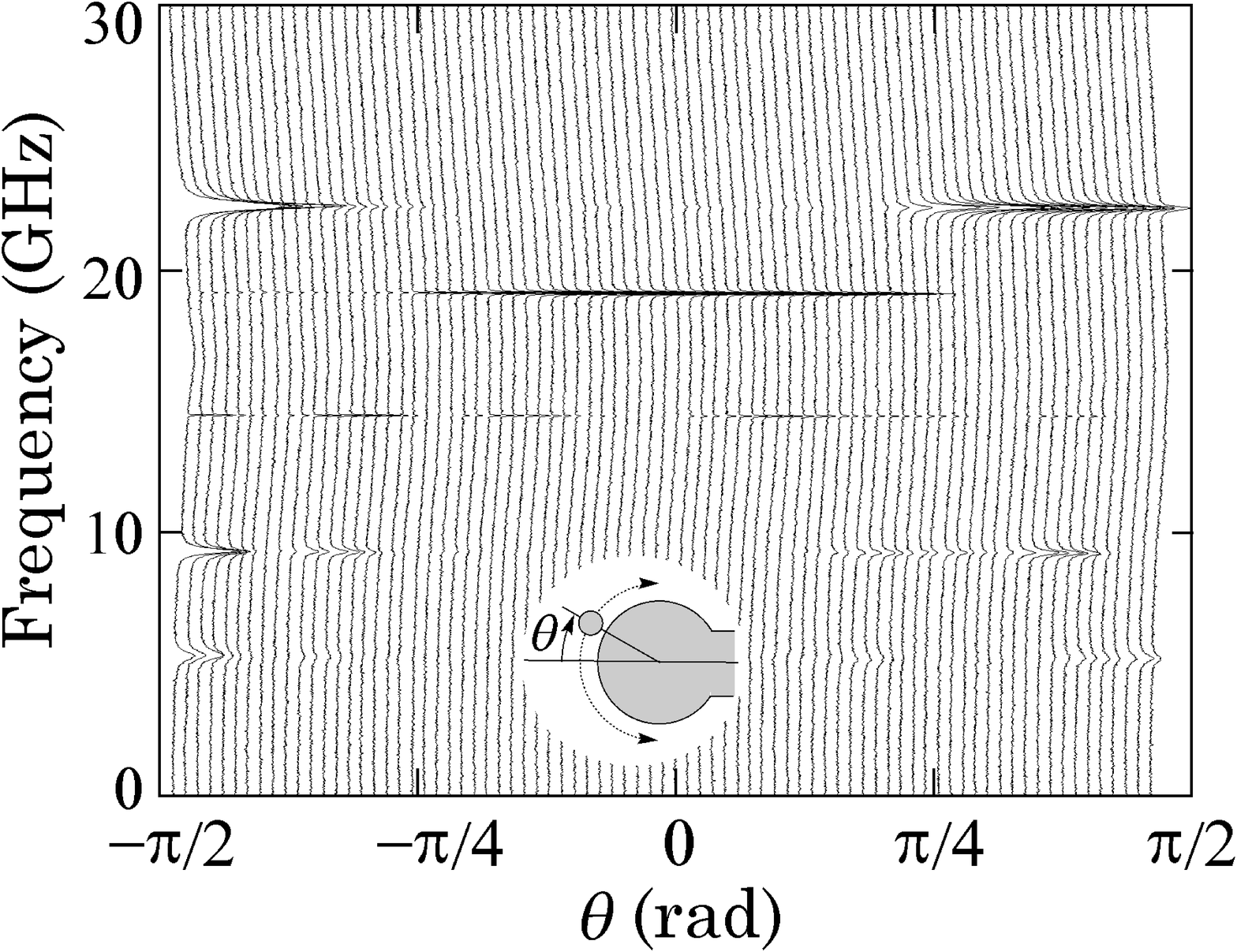}
  \caption{\small Circular taper mapping of a microtoroid. The spectra of the transmitted power are offset to the corresponding $\theta$ position. Inset : definition of $\theta$ coordinate.}
  \label{f:modestoroid}
\end{figure}

The WGMs of toroidal microcavities have the advantage to achieve smaller mode volumes than microspheres of the same outer diameter. A general analytical theory for their mode structure and resonance positions is not available, in particular because the $n$ and $\ell$ no longer exist while $m$ (angular momentum) and $q$ keep their significance. An experimental determination of their field distribution is highly desirable. Our new method can conveniently be adapted to monitor this field distribution. Because of the large curvature of the toroid's minor-diameter, a vertical scanning would result in a very large gap, and a vanishing signal. Therefore we have replaced the linear $z$ scanning by a circular $\theta$ scanning, intended to keep an almost constant coupling gap ($\theta$ is defined on the inset of Fig.~\ref{f:modestoroid}).  We have used rather thick microtoroids (minor diameter $d\approx 6.5\units{\upmu m}$), in order to observe several modes in a typical frequency range of $50\units{GHz}$. The result of this experiment is shown in Fig.~\ref{f:modestoroid}. It is displayed as a waterfall, where successive spectra are offset according to $\theta$, from $-\pi/2$ (below the toroid) to $\pi/2$ (above the toroid). This figure shows several modes with clearly distinct angular intensity distributions, and in particular a fundamental mode at a frequency offset of $19\units{GHz}$. One can notice that the resonances are broader and more pronounced on the left side of the figure because of an imperfect centering of the circular motion, leading to a smaller gap on this side than on the opposite one, but this does not prevent the WGMs' characterization.

To conclude, we have demonstrated a new and robust method to characterize the angular structure of WGMs of a microcavity. Based on the taper used for excitation, it eliminates steric problem arising for other methods. It allows to accurately position the coupling device at the equator location, thus optimizing the coupling to the most confined mode and canceling odd-modes coupling efficiency.

This work has been supported by C'Nano-Île de France and French embassy in Beijing. G.~L. acknowledges support from the Chinese Scholarship Council.


\end{document}